\documentclass[preprint,11pt, 3p,twocolumn]{elsarticle}

\usepackage{color,graphicx,epsfig}
\usepackage{ifpdf}
\usepackage{amsmath}
\usepackage{bm}
\usepackage{color}
\usepackage[english]{babel}
\usepackage{graphicx}%
\usepackage{amsfonts}%
\usepackage{amssymb}
\usepackage{braket}
\usepackage{hyperref}
\usepackage{enumerate}

\usepackage[english]{babel}
\usepackage{blindtext}

\definecolor{nicered}{rgb}{0.7,0.1,0.1}
\definecolor{nicegreen}{rgb}{0.1,0.5,0.1}
\hypersetup{colorlinks,citecolor= nicegreen,linkcolor= nicered}

\newcommand{\be}{\begin{equation}}
\newcommand{\ee}{\end{equation}}
\newcommand{\bea}{\begin{eqnarray}}
\newcommand{\eea}{\end{eqnarray}}

\definecolor{Red}{rgb}{1.,0.,0.}

\journal{Physics  Letters B}

\title{Higgs bosons with  large transverse momentum at the LHC}

\author[kit]{Kirill Kudashkin}
\ead{kirill.kudashkin@kit.edu}
\author[ippp]{Jonas M. Lindert}
\ead{jonas.m.lindert@durham.ac.uk}
\author[kit]{Kirill Melnikov}
\ead{kirill.melnikov@kit.edu}
\author[kit,ikp]{Christopher Wever}
\ead{christopher.wever@kit.edu}

\address[kit]{Institute for Theoretical Particle Physics (TTP), KIT, Karlsruhe, Germany}
\address[ippp]{Institute for Particle Physics Phenomenology, Durham University, 
Durham, DH1 3LE, UK}
\address[ikp]{Institut f\"ur Kernphysik (IKP), KIT, 76344 Eggenstein-Leopoldshafen, Germany}

\cortext[thanks]{Preprint number: IPPP/18/5, TTP18-03 }

\begin{document}

\begin{frontmatter}

\begin{abstract}
We compute the next-to-leading order QCD corrections to 
the production of Higgs bosons  with  large  transverse momentum $p_\perp \gg 2 m_t$ at the LHC. 
To accomplish  this, we combine the two-loop amplitudes for  processes 
$gg \to Hg$, $qg \to Hq$ and $q \bar q \to H g$, recently computed 
in the approximation  of nearly massless top quarks, with the  numerical 
calculation of the squared one-loop amplitudes for $gg \to Hgg$, $q g \to H q g$ 
and $q \bar q \to Hgg$ 
processes. The latter computation is   performed with  {\tt OpenLoops}.  
We find that the QCD corrections to the  Higgs transverse momentum 
distribution at very high $p_\perp$  are large  but quite  similar 
to  the QCD corrections obtained   for point-like $Hgg$ coupling. 
Our result removes one of the largest sources of theoretical 
uncertainty in the description of  high-$p_\perp$ Higgs boson  production 
and opens a way to use 
the  high-$p_\perp$  region to search 
for physics beyond the Standard Model. 
\end{abstract}


\end{frontmatter}


\section{Introduction}
\label{intro}

Detailed exploration of the Higgs boson is one of the central tasks of the particle 
physics program at the LHC.  Since the majority of the Higgs bosons  is  produced in 
the gluon fusion, it is only natural 
to study Higgs coupling to gluons as precisely as possible. 

Incidentally, the Higgs-gluon coupling is very interesting phenomenologically. Indeed, 
since the Higgs coupling to gluons is loop-induced, and since contributions 
of heavy particles whose masses are generated by the Higgs mechanism do not decouple, the $ggH$ 
interaction vertex becomes an intriguing probe of the TeV-scale physics.

In the Standard Model, the $ggH$ interaction vertex  is almost entirely
generated by the top quark loops 
and, since the top  Yukawa coupling in the Standard Model is fully determined by the top 
quark mass, it appears that the $ggH$ coupling is fully predictable. However, since 
the top Yukawa coupling is known experimentally to about $50$ percent from  $ttH$ production 
process \cite{Aad:2016zqi,Aaboud:2017rss},  it is still  
possible  that there is additional, point-like component of the $Hgg$ coupling that appears 
thanks to physics beyond the Standard Model (BSM). 

To describe this possibility, we consider 
the following modification of the top Yukawa part of the SM  Lagrangian
\be
\begin{split}
& \frac{m_t}{v} \bar t t H \to 
\\
& -\kappa_g \frac{\alpha_s}{12 \pi v} G_{\mu \nu}^{a} G^{\mu \nu, a} H + \kappa_t \frac{m_t}{v} \bar t t H.
\label{eq1}
\end{split}
\ee
The first term on the r.h.s. in Eq.~(\ref{eq1}) is the point-like contribution to the Higgs-gluon 
coupling  and the second is the modified top Yukawa coupling. 

What are the constraints on the anomalous couplings $\kappa_g$ and $\kappa_t$ from the 
Higgs production in gluon fusion?   Since the top quark contribution 
to  Higgs boson production in gluon fusion is well-described in  
the large-$m_t$ approximation,  the production cross section 
is proportional to the sum of the two couplings squared 
$\sigma_{gg \to H} \sim \alpha_s^2/v^2 ( \kappa_g + \kappa_t)^2$. 
Clearly, even if  the cross section  $\sigma_{gg \to H}$ is measured with 
absolute precision, we  cannot constrain  $\kappa_{g}$ and $\kappa_t$ separately but 
only their sum. 

To disentangle $\kappa_g$ and $\kappa_t$,  one has to go beyond  total cross section measurements.  
A useful and simple observable \cite{zupan} is the Higgs boson transverse momentum distribution. 
Indeed, if we assume that scale of New Physics\footnote{We refer to such a scale 
as $\Lambda_g$.} 
that generates  point-like $ggH$ coupling   proportional to $\kappa_g$, 
is much larger  than twice the top quark mass,  there 
exists a  range of  transverse momenta $ 2m_t \ll p_\perp \ll \Lambda_g$ such that 
the BSM  contribution to the Higgs-gluon vertex    
can still be treated as point-like, 
whereas the top quark contribution  starts being resolved. 
This feature can be illustrated by the following schematic formula
\be
\frac{{\rm d} \sigma_H}{{\rm d} p_\perp^2} \sim \frac{\sigma_0}{p_\perp^{2} } 
\left \{
\begin{array}{cc} 
 (\kappa_g + \kappa_t)^2 , &   p_\perp^2 < 4m_t^2, \\
 \left ( \kappa_g + 
\kappa_t \frac{4m_t^2}{p_\perp^2}  \right )^2,   & p_\perp^2 >4 m_t^2.
\end{array}
\right. 
\label{eq2}
\ee
This formula suggests that a measurement of the Higgs transverse momentum distribution 
in the two regions, $p_\perp \ll 2m_t$ and $p_\perp \gg 2 m_t$, allows for a separate determination 
of $\kappa_g$ and $\kappa_t$. 

There are quite a few obstacles to a practical realization of this program. 
First, assuming that 
$\kappa_t \sim 1$, $\kappa_g \sim 0.1$, the $p_\perp$ distribution 
of the Higgs boson is dominated by the Standard Model contribution 
until rather high values of the Higgs transverse momentum. 
Unfortunately, since  the cross section decreases quite  fast with $p_\perp$, 
we expect a relatively small number of events in the interesting 
transverse momentum region. 
For example, the  SM   cross section for producing  Higgs bosons with  transverse momenta
larger  than $ 450~{\rm GeV}$ is close to ${\cal O}(10~{\rm fb})$; therefore, even 
allowing for small deviations from the Standard Model, we estimate  that 
just a few hundred 
Higgs bosons have been produced in the  
high-$p_\perp$ region at the LHC so far.

Second,  even if Higgs bosons with high transverse momenta  are produced, 
identifying them through  
standard clean decay channels $H \to \gamma \gamma$ and $H \to 4~{\rm leptons}$ 
decreases the number of events 
because of the tiny branching fractions  of these  decay modes. In fact, the number 
of events is reduced to such an extent that, 
given  current  integrated luminosity, it  becomes impossible to observe them.

The third point concerns the quality 
of the theoretical description of the Higgs $p_\perp$ spectrum at high transverse momentum. 
As follows from Eq.~(\ref{eq2}), we require the  description of the spectrum 
in two regimes: a) $p_\perp < 2 m_t$, where 
the $ggH$ interaction is, effectively,
point-like and b) $p_\perp > 2 m_t$ where, in addition to the 
point-like interaction,  there is a ``resolved'', $p_\perp$-dependent 
component  due to  the top quark loop.   Theoretical description of 
Higgs boson production in gluon fusion, for a point-like gluon-Higgs  vertex, 
is extremely advanced. Indeed, the inclusive rate for gluon fusion 
Higgs production in this approximation is  
known to the astounding  N$^3$LO QCD accuracy
\cite{Anastasiou:2016cez},  and the Higgs $p_\perp$-distribution 
has been computed to NNLO QCD \cite{Boughezal:2015dra,Boughezal:2015aha, Chen:2014gva}.

In comparison, very little is known about gluon fusion {\it beyond} 
the point-like approximation for the $ggH$ interaction vertex 
which becomes of particular relevance at high $p_\perp$.  The corresponding 
cross section was computed at leading order  in perturbative QCD 
\cite{Ellis:1987xu} {\it thirty years ago} and only recently 
this result was extended to next-to-leading order in a situation 
when the mass of the quark that facilitates the Higgs-gluon interaction 
is much smaller than the Higgs boson mass and all other kinematic invariants in the 
problem \cite{Melnikov:2016qoc,Lindert:2017pky}.

We emphasize that even if the first and  second points 
that we mentioned earlier can be overcome, imprecise knowledge of the 
Standard Model contribution at high $p_\perp$ may be an obstacle 
for the determination of $\kappa_g$. Indeed,  since the two contributions 
to ${\rm d} \sigma/{\rm d} p_\perp^2$ at high 
$p_\perp$ may receive different radiative corrections,  lack of their knowledge 
may  affect   the interpretation of the result especially if relatively 
small  values of $\kappa_g$ are to be probed. 
Since for processes with gluons in the initial state large QCD corrections 
are typical,  one can expect large 
radiative corrections also for the resolved top-quark  loop at high $p_\perp$. 
Although the fact that radiative corrections are large is almost 
guaranteed, the important  question is  
by how much they differ if the high-$p_\perp$ tail  
of the Higgs transverse momentum distribution  is computed with 
the point-like or ``resolved''  Higgs-gluon  vertex. This is the question that 
we attempt to answer in this paper. 

It is clear that the low statistics issue, that was  mentioned in the 
 first and second points  above, can only be overcome by collecting  
higher integrated luminosity;  luckily, the LHC will continue 
doing that. However, it should be possible, already now,  to perform  relevant 
measurements in the high-$p_\perp$ region if one does not lose 
so much statistics by insisting that  the produced Higgs bosons should 
decay into clean final states. 
Interestingly,  it appears to be 
possible to do that. Indeed, in contrast to low-$p_\perp$ Higgs production, 
at high-$p_\perp$ one can  identify the Higgs boson 
through its decays to $H \to b \bar b$ using the  boosted techniques 
\cite{gavin}
and 
to distinguish hadronically-decaying Higgs bosons from  large QCD backgrounds.  
In fact, the CMS collaboration has recently 
presented results of the very 
first analysis \cite{CMS:2017cbv} performed along these lines, 
where Higgs boson production with $p_\perp > 450$~GeV, was observed. 
Although  the result for the Higgs production cross section with 
$p_\perp > 450$~GeV obtained in \cite{CMS:2017cbv}
is rather imprecise,   
forthcoming improvements with higher luminosity and better 
analysis technique are to be expected. 

The third point mentioned above  is an important issue. 
Indeed, since the Standard Model production 
of a  Higgs boson at  high $p_\perp$ involves ``resolved'' 
top quark loops,  computing 
next-to-leading order QCD corrections to this process requires 
dealing with  two-loop  four-point functions with internal  (top quark) and 
external (Higgs boson) massive particles.  
The relevant two-loop Feynman 
integrals with the full dependence on $m_t$ and $m_H$ are still not 
available,\footnote{We note that 
planar two-loop integrals for this process 
were computed in Ref.~\cite{hjalte}}
so that  the NLO QCD computation can not be performed.
Thus, in the literature various approximations have been performed
both for inclusive Higgs production~\cite{approxrefinc}
and also for finite Higgs $p_\perp$ \cite{approxrefexl,Neumann:2016dny}.
However,   recently, 
 the two-loop amplitudes for the production of the Higgs boson at high-$p_\perp$
were computed
\cite{Kudashkin:2017skd}. This result enables 
calculation of the Higgs boson transverse momentum  distribution for $p_\perp > 2m_t$ 
at NLO QCD, that we report in this Letter.

The rest of the paper is organized as follows.  In the next 
Section, we provide a short summary of theoretical methods 
used for the calculation of two-loop virtual and real emission corrections. 
Phenomenological results are reported in Section~\ref{results}. 
 We conclude in Section~\ref{concl}. 
 
\section{Computational Setup}
\label{framework}

We begin with the discussion of the computation of the two-loop QCD 
amplitudes  for producing the Higgs boson  with  large transverse momentum 
in proton collisions \cite{Kudashkin:2017skd}.   
There are four partonic processes that  contribute;  they are 
$ gg \to Hg $, $ g\bar{q} \to H\bar{q} $, $ gq \to Hq $ 
and $ q\bar{q} \to Hg $.  We systematically neglect the Higgs boson coupling to 
light quarks; therefore,  all contributions to scattering amplitudes are mediated 
by the top quark loops and are proportional to the top quark Yukawa coupling.

In principle, scattering amplitudes for  Higgs boson production with non-vanishing  
transverse  momentum depend 
on the Higgs boson mass, the top quark mass and two  Mandelstam invariants $s$ and $t$. 
Computation of two-loop Feynman integrals  that depend on such a large number of parameters 
and, moreover, contain internal massive lines, is, currently not feasible. 
However, since we are interested in describing production 
of the Higgs boson with  {\it high} transverse momentum, 
we can construct an expansion of the scattering 
amplitude in $m_i^2/s$ and  $m_i^2/p_\perp^2$, where $m_i^2 \in \{ m_H^2, m_t^2\}$.  
Additionally, as 
$m_H^2/(2m_t)^2 \sim 0.1$,  it is motivated
to neglect the Higgs boson mass
compared to the top quark mass in the computation. 

It is however non-trivial to construct such an expansion in the Higgs and top quark masses. 
Indeed, in contrast to an opposite 
kinematic limit, $p_\perp \ll m_t$, where expansion of scattering amplitudes can be performed 
at the level of Feynman integrals in momentum space  using the large-mass expansion algorithms
\cite{Smirnov:2013}, no momentum-space 
algorithms exist for an expansion in the small  quark mass.  
For this reason, 
we have opted for a different method \cite{Mueller:2015lrx,Melnikov:2016qoc}. 
The idea is to first derive 
differential equations for master integrals\footnote{The required algebraic 
reduction to master integrals is highly non-trivial; for this, we have used 
results obtained in an earlier collaboration with L.~Tancredi in Ref.~\cite{Melnikov:2016qoc}.}
that are {\it exact} in all kinematic parameters and then develop a systematic expansion of these 
differential equations in the mass of the top quark and the Higgs boson.  Since the differential 
equations contain all the information about the possible  singularities of the solutions, 
we can construct the expansion of the solutions in the limit of small Higgs and top 
masses.

\begin{figure*}[htb]
\centering
\hspace{-0.9cm} \includegraphics[width=.59\textwidth]{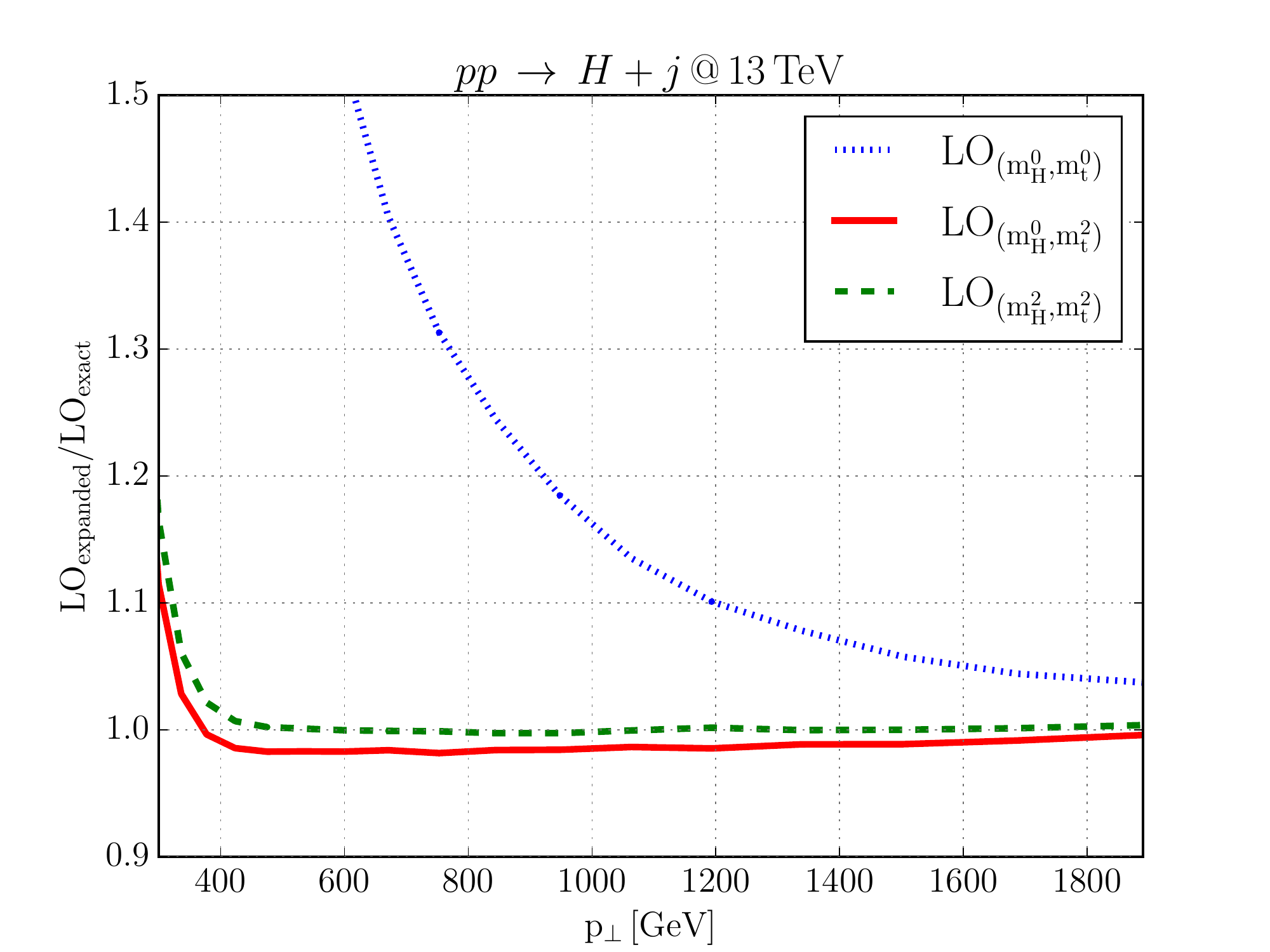} 
\caption{Ratio of approximate to exact  
leading order cross sections. By retaining   ${\cal O}(m_t^2/p_\perp^2)$ 
corrections in scattering 
amplitudes (red line), we obtain an excellent approximation to the exact LO result. The notation 
$\text{m}_{\text{t}}^0$ and $\text{m}_{\text{t}}^2$ 
in the legend of the plot refers to the leading and next-to-leading power expansion of the amplitude in $m_t^2$, 
not including the overall $m_t^2$ that arises from the Yukawa coupling and the helicity flip.
}
\label{fig:LO}
\end{figure*}

We have applied this method to compute all the master integrals relevant for 
the Higgs+jet production 
amplitudes \cite{Kudashkin:2017skd}.   When computing these amplitudes, 
we have retained first sub-leading terms in the $m_t^2/p_t^2$-expansion but 
we have set the mass of the Higgs 
boson to zero.\footnote{Setting $m_H$ to zero is possible 
since the dependence of all  amplitudes on the Higgs boson mass is analytic.}     
It turned out that by 
keeping sub-leading terms in the $m_t^2/p_\perp^2$-expansion in the amplitude  we can
 significantly extend  the applicability range of the computation. 
To illustrate this, in Fig.~\ref{fig:LO},  we compare  the exact leading order $p_\perp$ distribution 
of the Higgs boson with three expansions. We see that the result for amplitude expanded to 
${\cal O}(m_H^{0},  m_t^{2})$ terms tracks the leading order amplitude at the level of few percent
all the way down to the top quark threshold; on the contrary, if  the sub-leading top quark 
mass terms are not retained, the expanded and exact cross sections  
have ${\cal O}(20\%)$  difference  at $p_\perp \sim  800~{\rm GeV}$. 
Even higher terms in the $m_t^2/p_\perp^2$-expansion do not further improve the agreement
at the scale of Fig.~\ref{fig:LO}.
Yet, keeping also subleading terms in the $m_H^2$ expansion can further improve the agreement with the
exact result. For illustration, in Fig.~\ref{fig:LO} we show the result for the amplitude expanded up to 
${\cal O}(m_H^{2},  m_t^{2})$, which above threshold agrees at the permil level with the exact result.
Including even higher expansions does not yield further improvement visible at the scale of Fig.~\ref{fig:LO}.

In order to produce physical results for  Higgs boson  production with non-vanishing 
transverse momentum, we need to combine 
the above discussed virtual corrections with the corresponding real corrections, e.g. $gg \to H+gg$, $qg \to Hq+g$ etc,
 that describe inelastic processes. 
Computation of one-loop scattering amplitudes for these inelastic processes is non-trivial; 
it requires evaluation of five-point Feynman integrals  with massive internal particles.
Nevertheless,  these amplitudes are known analytically since quite some time~\cite{DelDuca}.\footnote{These 
amplitudes were recently re-evaluated in Ref.~\cite{Neumann:2016dny}.}

In this Letter we follow an approach, based on the 
automated numerical computation of one-loop scattering amplitudes developed in recent years.  
One such approach, known as 
OpenLoops~\cite{Cascioli:2011va}, employs a hybrid tree-loop recursion.
Its implementation in the {\tt OpenLoops} program is publicly available~\cite{openloops,openloopstwo}. 
This program
has been applied to compute one-loop QCD and electroweak corrections to
a multitude of complicated multi-leg scattering processes 
(see e.g. Refs. \cite{olrefs,Jezo:2016ujg})  and for  the real-virtual 
contributions in NNLO computations (see e.g. Ref.~\cite{Cascioli:2014yka}).  

For these applications in NNLO calculations and  for computing NLO corrections 
to loop-induced processes, such as the one discussed here,
the corresponding one-loop real contributions need to be computed in 
kinematic regions where one of the external partons becomes soft or 
collinear to other partons. A reliable evaluation of the one-loop scattering 
amplitudes in such kinematic regions is non-trivial, but {\tt OpenLoops} appears to be  
perfectly capable of dealing  with this challenge
thanks to the numerical stability of the employed algorithms. A major 
element of this stability originates from the employed tensor integral reduction 
library {\tt COLLIER}~\cite{collier}.

{\renewcommand{\arraystretch}{1.5}
\begin{table*}[t]
  \vspace*{0.3ex}
  \begin{center}
\begin{small}
    \begin{tabular}{c|ccc|ccc}
& 
  LO$_{\rm HEFT}$ [fb]   &     NLO$_{\rm HEFT}$ [fb]   &   $K$   &
  LO  [fb] & NLO [fb]  &   $K$  \\
\hline
$p_\perp$$>400~$GeV  & 
  $33.8^{+44\%}_{-29\%}$  &   $61.4^{+20\%}_{-19\%}$      &  $1.82$  &
  $12.4^{+44\%}_{-29\%}$  &   $23.6^{+24\%}_{-21\%}$ & $1.90$                                 
\\ 
$p_\perp$$>450~$GeV  & 
  $22. 0^{+45\%}_{-29\%}$  &   $39.9^{+20\%}_{-19\%}$      &  $1.81$  &
  $6. 75^{+45\%}_{-29\%}$  &   $12.9^{+24\%}_{-21\%}$    &     $1.91$                        
\\ 
$p_\perp$$>500~$GeV  & 
  $14.7^{+44\%}_{-28\%}$  &   $26.7^{+20\%}_{-19\%}$      &  $1.81$  &
  $ 3.80^{+45\%}_{-29\%}$ &   $7.28^{+24\%}_{-21\%}$    &     $1.91$                        
\\ 
$p_\perp$$>1000~$GeV & 
  $0.628^{+46\%}_{-30\%}$   &   $1.14^{+21\%}_{-19\%}$       &  $1.81$  &
  $0.0417^{+47\%}_{-30\%}$   &   $0.0797^{+24\%}_{-21\%}$       & $1.91$                                 
    \end{tabular}
\end{small}
  \end{center}
  \caption{
Inclusive cross sections and $K$-factors 
for $pp\to H$+jet at $\sqrt{S}$=13\,TeV in the SM and 
in the infinite top mass approximation with different lower cuts on the 
Higgs boson transverse momentum  $p_\perp$. We estimate the theoretical 
uncertainty in the predicted cross section by changing renormalization and 
factorization scales by a factor of two around the central 
value in Eq.(\ref{eqscale}).    We define 
the $K$-factors as $\sigma_{\rm NLO}/\sigma_{\rm LO}$. The results for $K$-factors 
in the Table are computed for the central value of the renormalization scale.  
 See text for details. 
}
\label{tab:results}
\end{table*}
}

All virtual and real amplitudes have been implemented in the {\tt POWHEG-BOX}~\cite{powheg}, where 
infra-red  singularities are regularized using 
FKS subtraction~\cite{Frixione:1995ms}. All {\tt OpenLoops} 
amplitudes 
are accessible via  a process-independent  interface developed in Ref.~\cite{Jezo:2016ujg}. The implementation
within the {\tt POWHEG-BOX} will allow for an easy  matching of the fixed-order results presented here with
parton showers at NLO.

\section{Results}
\label{results}

In this Section, we present the results of our computation of the NLO QCD corrections 
to Higgs boson  production at high $p_\perp$.  
We consider proton-proton collisions at the LHC with the center of mass energy $13~{\rm TeV}$. 
The Higgs boson mass\footnote{Although the Higgs boson mass is ignored in the two-loop virtual 
amplitude,  it is retained  
in the computation of the real emission contribution to the 
transverse momentum distribution.} 
and the top quark mass\footnote{We renormalize 
the top quark mass in the pole scheme.}  are taken to be  
$m_H = 125~{\rm GeV}$ and  $m_t=173.2~$GeV,  respectively.
We employ the five-flavor  scheme and consider the bottom quark as massless parton in 
the proton.  We use the NNPDF3.0 set of parton distribution 
functions \cite{Ball:2014uwa} at the respective perturbative
order and employ the strong coupling constant $\alpha_s$ that is provided with these  
PDF sets.  We choose renormalization and factorization scales to be equal and take
as the central value
\begin{equation}
\mu_0 = \frac{H_T}{2}, \;\;\; H_T = \sqrt{m_H^2 + p_\perp^2} + \sum_{j}^{}  p_{\perp,j}\,,
\label{eqscale}
\end{equation}
where the sum runs over all partons in the final state. 
We note that at  large Higgs boson transverse momentum, the  
scale simplifies to  $\mu_0 = H_T/2 \approx p_\perp$. Theoretical uncertainties 
are estimated by varying  the renormalization and factorization scales $\mu$ by a factor of two 
around the central value.  
Finally, we note that we use both the leading 
order  and the  real-emission amplitudes for  Higgs boson production with 
non-vanishing transverse momentum keeping full dependence on the top quark 
and Higgs boson 
masses   and we only use the expansion in the top quark mass and the Higgs
boson mass in the finite remainder of the two-loop amplitude.

\begin{figure*}[htb]
\centering
\includegraphics[width=.65\textwidth]{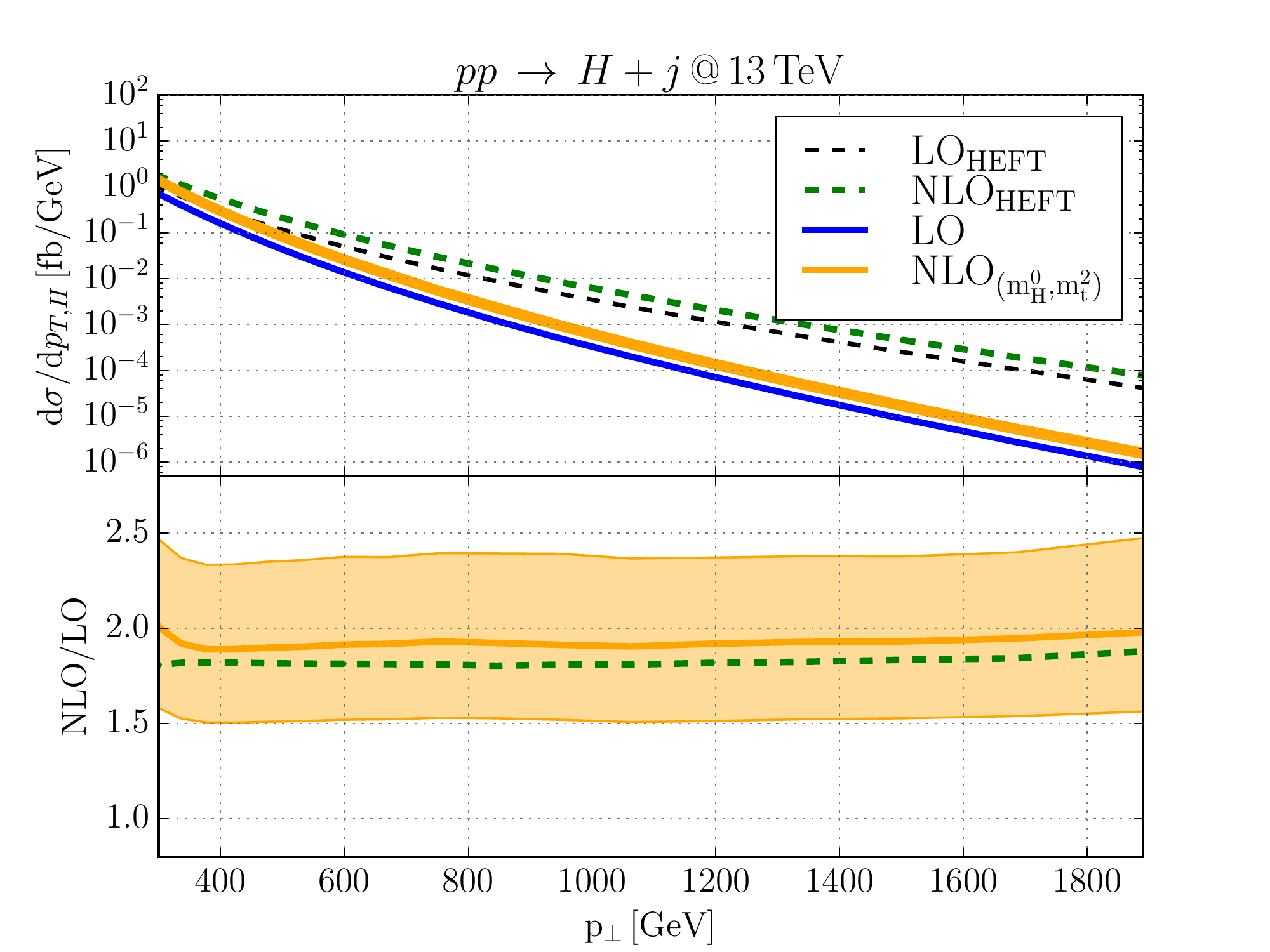}
\caption{Transverse momentum distribution of the Higgs boson at the LHC with $\sqrt{s}$=13\,TeV. 
The upper panel shows absolute predictions at LO and NLO in the full SM and in the 
infinite top-mass approximation (HEFT). The lower panel shows 
respective NLO/LO correction factors. The bands indicate theoretical errors 
of the full SM result due to scale variation. 
}
\label{fig:NLO}
\end{figure*}

The results of the computation are presented in  Table~\ref{tab:results} where we show 
the inclusive cross sections at LO and NLO together  
with the corresponding NLO/LO correction factors for different values of the lower cut 
on the Higgs transverse momentum.  The inclusive cross sections are computed for both 
the point-like Higgs-gluon coupling, obtained by integrating out the top quark, 
and for the physical Higgs-gluon coupling with a proper dependence on $m_t$.  We will 
refer to the two cases as HEFT and SM, respectively. 
Although the differences between HEFT and SM production cross sections grow dramatically 
with the increase of the $p_\perp$-cut, the radiative corrections change both cross sections 
by a similar amount. Indeed, the ratio of $K$-factors\footnote{The $K$-factors 
are defined as ${\rm d}\sigma_{\rm NLO}/{\rm d}\sigma_{\rm LO}$.}
for the 
$p_\perp = 400~{\rm GeV}$ cut  and the central scale $\mu_0$ 
is  $K_{\rm SM}/K_{\rm HEFT} =1.04$ and 
the ratio of $K$-factors for the $p_\perp = 1000~{\rm GeV}$ cut  is 
$K_{\rm SM}/K_{\rm HEFT} =1.06$.  Note that the $K$-factor themselves are close 
to $1.9$, almost independent of the $p_\perp$-cut. 
Uncertainties due to scale variations are reduced from about $40\%$ at LO to the level
of $20\%$ at NLO, both for a point-like Higgs-gluon coupling and in the full SM. These
uncertainties are insensitive to the $p_\perp$-cut.

Finally, the Higgs boson transverse momentum distribution 
for $p_\perp > 300~$GeV is shown in Fig.~\ref{fig:NLO}.  The results 
shown there confirm what is already seen in Table~\ref{tab:results} -- 
both the SM and the HEFT $K$-factors are flat over the entire range of 
$p_\perp$.  For the central scale 
$\mu = \mu_0$  Eq.~(\ref{eqscale}) 
the differences between the two $K$-factors  is about  five percent. 
The scale dependence of HEFT and SM results are also similar. The residual theoretical 
uncertainty related to  perturbative QCD computations remains at the level of twenty 
percent, as estimated from the 
scale variation. Such an uncertainty is typical for NLO QCD theoretical description 
of  many observables related to Higgs boson production in gluon fusion. 

Another source of uncertainties is related to the choice of the renormalization-scheme 
of the top mass. Since the amplitude is proportional to the squared top mass, the 
differential cross section scales as the fourth power $d\sigma \sim m_t^4$, if we 
neglect suppressed terms in $m_t^2/p_{\perp}^2$ and the logarithms of $m_t^2/p_{\perp}^2$. At LO in 
perturbation theory, a different choice of the top-mass scheme corresponds to 
changing numerically the input value of the top mass. If we choose instead 
the $\overline{\rm MS}$ top mass value\footnote{We calculated this value using the program RunDec \cite{rundec} 
with the input value $ m_t^{\overline{ \rm MS}} ( m_t^{\overline{ \rm MS}} )~=~166~{\rm GeV}$.} 
of $m_t^{\overline{\rm MS}}(p_{\perp}\approx 400\, \rm{GeV})\approx 157\, \rm{GeV}$, we 
would find a decrease of the LO cross section by 
about $d\sigma_{\rm{LO}}^{\overline{ \rm MS}}/d\sigma_{\rm{LO}}^{\rm pole} \sim (157/173)^4\sim 0.68$. 
At NLO one needs to additionally take into account the $\alpha_s$ corrections that relate the on-shell and $\overline{ \rm MS}$ 
top mass values. These corrections will compensate the numerical change caused by 
changing $m_t=m_t^{\overline{ \rm MS}}$ to $m_t=m_t^{\rm pole}$ in the NLO amplitudes and as a result 
the scheme dependence at NLO is reduced. Thus, we expect the scheme dependence at NLO to be subleading with respect to the scale uncertainties.

Further improvements in theory predictions are only possible if the proximity of the 
HEFT and SM $K$-factors is taken seriously and postulated to occur even at higher orders. 
In this case, one will have to re-weight the existing HEFT $H+j$ 
computations \cite{Boughezal:2015dra,Boughezal:2015aha,Chen:2014gva} 
with the exact leading order cross section 
for producing the Higgs boson with high $p_\perp$. In fact, such a reweighting 
can now be also performed at the NLO level. 

\section{Conclusions} 
\label{concl} 

We presented the  NLO QCD corrections to the Higgs boson 
transverse momentum distribution  at very large $p_\perp$ values.  
To compute them, we employed the  recent calculation  of the 
two-loop  scattering amplitudes for all relevant partonic channels \cite{Kudashkin:2017skd}
 where an expansion in $m_t/p_\perp$ 
was performed.  The real emission 
corrections where computed with the {\tt Openloops} \cite{Cascioli:2011va} program. 
We have found that the QCD corrections 
to the Higgs boson transverse momentum distribution increase the leading order result 
by almost a factor of two. However,  their magnitude appears to be quite 
similar to the QCD corrections computed in the approximation of a  point-like 
Higgs-gluon vertex; the difference of the two result is close to five percent. 
Our computation removes the major theoretical uncertainty in the description of the Higgs 
boson  transverse momentum  distribution at high $p_\perp$ and opens a way to a refined 
analysis of the sensitivity of this observable to BSM contributions using existing 
 \cite{CMS:2017cbv} and forthcoming experimental measurements.

\vspace*{0.4cm}
{\bf Acknowledgments}
We thank Fabrizio Caola and Raoul R\"{o}ntsch for useful conversations. 
The research of K.M. was supported by the German Federal Ministry for %
Education and Research (BMBF) under %
grant 05H15VKCCA. The research of K.K. is supported by the DFG-funded Doctoral School KSETA %
(Karlsruhe School of Elementary Particle and Astroparticle Physics). %


\end{document}